\documentclass[runningheads]{llncs}

\usepackage[table,dvipsnames]{xcolor}

\usepackage{graphicx}
\usepackage{amsmath}
\usepackage{amssymb}
\usepackage{booktabs}
\usepackage{bm}
\usepackage{mathtools}
\usepackage{algpseudocode}
\usepackage{algorithm}
\usepackage{siunitx}
\usepackage{multirow}
\usepackage{enumitem}
\usepackage{caption}
\usepackage{wrapfig}
\usepackage{diagbox}
\usepackage{sidecap}
\usepackage{adjustbox}

\usepackage[breaklinks,colorlinks,citecolor=green]{hyperref}

\graphicspath{ {images/} }

\algnewcommand\algorithmiclocalize{\textbf{Localize neuron and update color model:}}
\algnewcommand\Localize{\item[\algorithmiclocalize]}
\algnewcommand\algorithmiccross{\textbf{Cross-section plane determination:}}
\algnewcommand\Cross{\item[\algorithmiccross]}
\algnewcommand\algorithmicbranchc{\textbf{Bifurcation candidates detection:}}
\algnewcommand\Branchc{\item[\algorithmicbranchc]}

\newcommand{\fft}[1]{\bm{\mathbf{\hat{#1}}}}
\newcommand{\norm}[1]{\vert \vert{#1}\vert \vert^2}
\newcommand{\normd}[1]{\vert \vert{#1}\vert \vert^2_2}
\newcommand{\diag}[1]{\text{diag}}
\newcommand{\conj}[1]{\text{conj}}

\makeatletter

\DeclareMathOperator*{\argmin}{arg\,min}

\usepackage[capitalize]{cleveref}
\crefname{section}{Sec.}{Secs.}
\Crefname{section}{Section}{Sections}
\Crefname{table}{Table}{Tables}
\crefname{table}{Tab.}{Tabs.}
\newcommand*\samethanks[1][\value{footnote}]{\footnotemark[#1]}

\begin{document}
\title{Online Multi-spectral Neuron Tracing}
\author{Bin Duan$^{1}$\thanks{Equal contribution. \;$^\dag$ Corresponding author.}\; Yuzhang Shang$^1$\samethanks\; Dawen Cai$^{2,3,4}$\; Yan Yan$^{1\dag}$}
\institute{
$^1$Department of Computer Science, Illinois Institute of Technology\\
$^2$Biophysics, $^3$Neuroscience Graduate Program, University of Michigan\\
$^4$Cell and Developmental Biology, Michigan Medicine\
\url{https://github.com/tuffr5/online-tracing}
}

\authorrunning{Bin Duan, Yuzhang Shang, Dawen Cai, and Yan Yan}
\titlerunning{Online Multi-spectral Neuron Tracing}

\maketitle              

\begin{abstract}
In this paper, we propose an online multi-spectral neuron tracing method with uniquely designed modules, where no offline training are required. Our method is trained online to update our enhanced discriminative correlation filter to conglutinate the tracing process. This distinctive offline-training-free schema differentiates us from other training-dependent tracing approaches like deep learning methods since no annotation is needed for our method. Besides, compared to other tracing methods requiring complicated set-up such as for clustering and graph multi-cut, our approach is much easier to be applied to new images. In fact, it only needs a starting bounding box of the tracing neuron, significantly reducing users' configuration effort. Our extensive experiments show that our training-free and easy-configured methodology allows fast and accurate neuron reconstructions in multi-spectral images.

\end{abstract}
\section{Introduction}

Neuron tracing is to reconstruct neuron structure in 3D image stack, capturing the trajectory of neurons traversing in the 3D volume, which is an essential step in neuroscience~\cite{livet2007transgenic,cai2013improved,gong2016high,januszewski2018high,abdolhoseini2019segmentation}. Recent advances in imaging and genetic techniques have enabled the capture of ultra-dense multi-channel a.k.a. multi-spectral images~\cite{livet2007transgenic,cai2013improved}. However, such high density can lead to severe cross-talk, i.e. overlapping among different neurons. Besides, the large number of spectral channels and imaging noise easily cause color drift, which largely degrades the interpretability of the neuron colors, making neuron tracing a challenging problem. 

Established methods such as active contour~\cite{schmitt2004new,chothani2011automated,wang2011broadly}, hierarchical path pruning~\cite{peng2011automatic,xiao2013app2} and graph multi-cut~\cite{quan2016neurogps,li2019precise} require additional set-up where they firstly process the images to obtain a rough neuron reconstruction, and then the tracing problem downgrades into refining the rough reconstruction into the most biologically persuasive counterpart~\cite{li2019precise,matejek2019biologically}. Other methods such as principal curve~\cite{bas2011principal}, template matching~\cite{zhao2011automated}, seed growing~\cite{choromanska2012automatic}, clustering~\cite{sumbul2016automated,duan2021unsupervised}, more or less, make assumptions for neurons' shape or appearance, which can be easily jeopardized by the heterogeneity of neurons, especially for multi-spectral images. Recent deep learning methods~\cite{jin2019shutu,huang2021automated,januszewski2018high,chen2021deep,wang2019multiscale,matejek2019biologically,sheridan2023local}, particularly for Electron Microscopy (EM) images~\cite{januszewski2018high,matejek2019biologically,sheridan2023local} and single-channel Light Microscopy (LM) images~\cite{jin2019shutu,huang2021automated,chen2021deep}, have achieved great success in neuron reconstruction. However, it takes enormous annotation to obtain such a well-trained model, and these methods usually perform suboptimal when handling multi-spectral images.

\begin{figure}[!t] \small
    \centering
    \includegraphics[width=\linewidth]{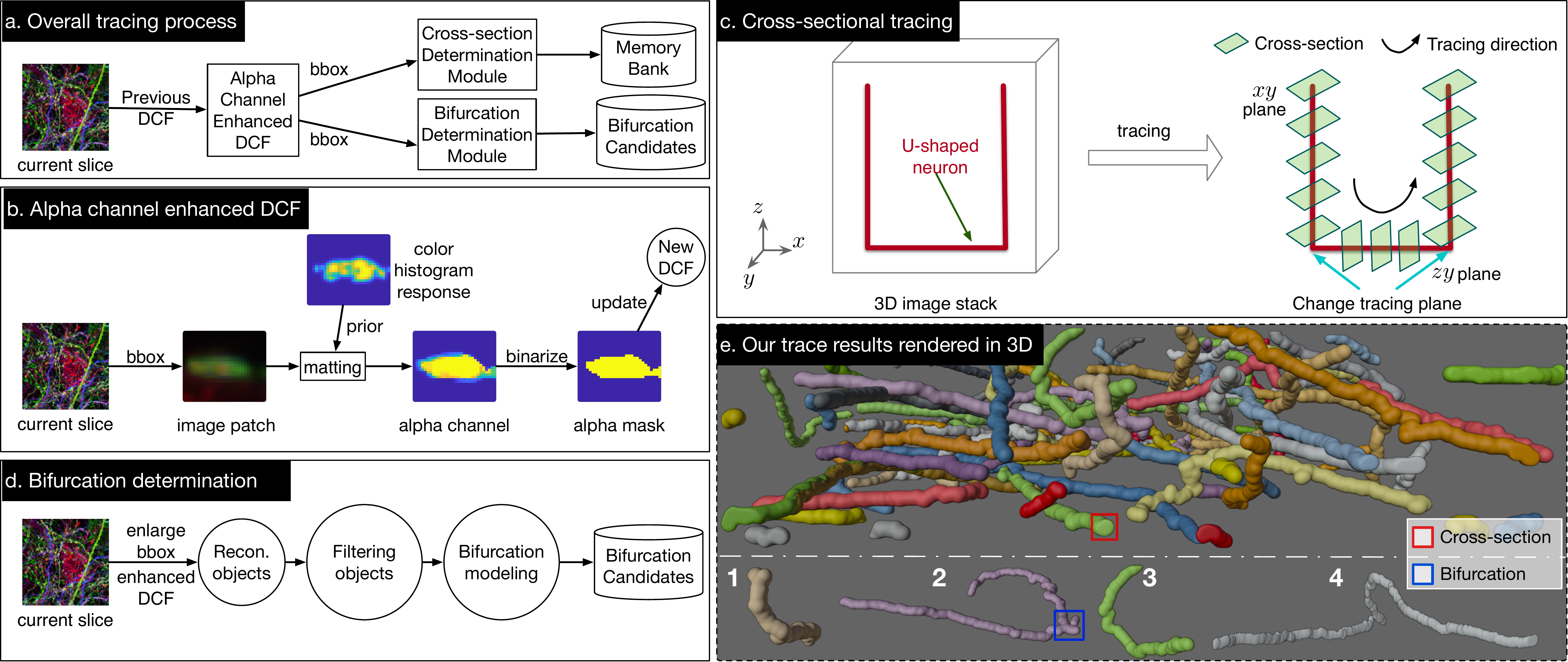}
    \captionsetup{skip=0pt,font=small}
    \caption{\textbf{Overview of the components of our proposed online multi-spectral neuron tracing.} Details of each component can be found in the Methodology section.}
    \label{fig:intro}
\end{figure}
Different from the aforementioned methods, in this paper, we propose a novel methodology, namely online neuron tracing for multi-spectral images. Given only the starting bounding box (bbox), our method learns online and maintains a global color model of the traced neuron so that it can deal with shape and appearance changes. Owing to both the online-learning property and straightforward set-up, our method is much easier to be applied to new images, compared to offline trained methods such as~\cite{januszewski2018high,matejek2019biologically,sheridan2023local,jin2019shutu}, as well as methods required complicated processing such as in~\cite{quan2016neurogps,li2019precise}. Moreover, to tackle difficult cases in neuron tracing, we propose two uniquely designed modules, i.e., cross-section and bifurcation determination modules, as illustrated in~\cref{fig:intro}.

Overall, (1) we introduce our xBTracer, the first online neuron tracing work, which is different from previous neuron tracing methodologies; (2) Our tracer can be executed given only the input bounding box and is training-free. So it is easy for neuroscientists to operate, compared to the complex set-up in previous methods; (3) Benefited from our uniquely designed modules, our tracer outperforms other methods by a large margin, verifying the effectiveness of our method; (4) Additional experiment show that our method can be generalized to other single-channel imaging modalities such as fMOST~\cite{gong2016high} data.
\section{Online Multi-spectral Neuron Tracer}
\noindent\textbf{Preliminary of Discriminative Correlation Filter (DCF).} 
Given a multi-spectral image patch $\mathbf{x} \in \mathcal{R}^{m\times n\times N_c}$, where $N_c$ is the channel number, and $m, n$ are height and width of the image patch. Following~\cite{bolme2010visual,lukezic2017discriminative}, the optimal filter $\mathbf{w}$ is found by minimizing the distance between the channel-wise feature filter correlation output and the ideal response $\mathbf{y}$ as,
\begin{equation}
    \argmin_\mathbf{w} \sum_{c=1}^{N_c} (\frac{1}{2} \norm{\diag{} (\fft{x}_c) \fft{w}_c^\top\!\!- \fft{y}_c} 
    + \frac{\gamma}{2} \norm{\mathbf{w}_c}).
    \label{eq:optimization}
\end{equation}
Here, $\fft{x}=\sqrt{D}\mathbf{Fx}$ is the Fourier transform, $\mathbf{F}$ is the discrete transform matrix and $D=m {\times} n$ is the length of $\mathbf{x_c}$, while $\diag{}(\cdot)$ being a $D \times D$ diagonal matrix. Build upon this naive DCF, we show that our enhanced DCF enables more compact neuron reconstruction, which is proven to be crucial for neuron traing.

\begin{figure}[!t]
    \centering
    \includegraphics[width=\linewidth]{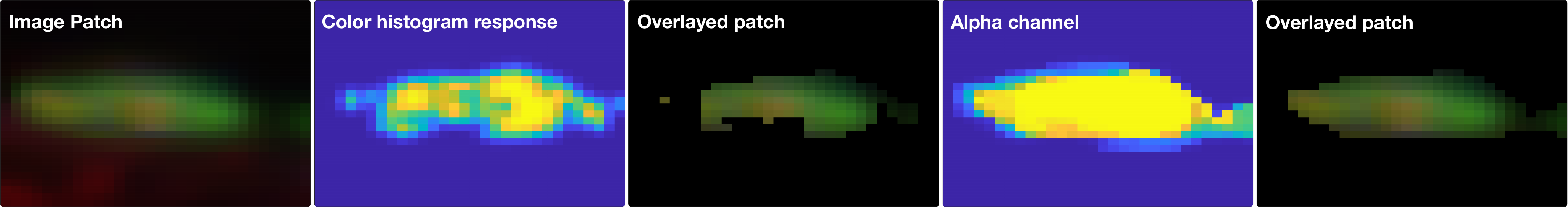}
    \captionsetup{skip=1pt,font=small}
    \caption{\textbf{Our alpha channel enables more compact reconstruction}. The overlayed patches are obtained by applying the corresponding mask to the same image patch.}
    \label{fig:alpha}
\end{figure}
\subsection{Alpha Channel Enhanced DCF}
\label{sec:aplha}
We first use previously learned DCF to locate bbox of the neuron in current slice to extract local image patch $\mathbf{x}$ to update a new DCF. Meanwhile, a global color model $\mathbf{g=\{g^f, g^b}\}$ is maintained for foreground and background color histograms from each bbox. Due to neuron's varieties, as shown in \cref{fig:alpha}, only using response of $\mathbf{g}$ leads into an incomplete reconstruction, easily distracting the tracing process. To obtain better reconstructions, we tackle this problem using image matting technique~\cite{Chen_2012KM,levin2007closed}. Give the prior, i.e. color histogram response $\mathbf{m}$ obtained from $\mathbf{g}$, we construct a feature vector $\mathbf{v}=(c_1,\cdots,c_{N_c},x,y)$ for each pixel, where $c_1,\cdots,c_{N_c}$ are the normalized intensity of each channel and $x,y$ are the normalized spatial coordinates. Thus, a matting Laplacian matrix $\mathbf{L}$ is formed by $\mathbf{v}$ and is then solved by the following quadratic problem using~\cite{barrett1994templates} as
\begin{equation}
    (\mathbf{L}+\lambda \diag{}(\mathbf{m}))\bm{\alpha}=\lambda \mathbf{m},
\end{equation}
where $\lambda$ controls the confidence of the prior $\mathbf{m}$. Next, we use Alternating Direction Method of Multipliers~\cite{boyd2011distributed} to optimize Eq.~\eqref{eq:optimization} as in the form of
\begin{equation}
     \mathbf{\Psi}(\mathbf{w}, \fft{f}, \fft{\zeta})=
    \frac{1}{2} \norm{\diag{} (\fft{x}) \fft{f}^\top\!\!- \fft{y}} 
    + \frac{\gamma}{2} \norm{\mathbf{w}_\alpha} + \fft{\zeta}^\top (\fft{f} - \fft{w}_\alpha) 
     + \frac{\mu}{2}\norm{\fft{f} - \fft{w}_\alpha},
\end{equation}
such that $\mathbf{f}-\diag{}(\bm{\alpha})\mathbf{w}=\mathbf{0}$. Here, binarized alpha mask $\bm{\alpha}$ serves as a constraint for the entire optimization, where $\mathbf{w}_\alpha \equiv \bm{\alpha} \circ \mathbf{w}= \diag{}(\bm{\alpha})\mathbf{w}$ for simplicity, $\bm{\alpha}$ and $\mathbf{w}$ are column vectors, and $\bm{\zeta}$ is the Lagrange multiplier. This is achieved by

\noindent\textbf{Solving} $\mathbf{\fft{f}^\star}$:\quad $\fft{f} = \frac{\fft{x} \circ \conj{}(\fft{y})+ \mu \fft{w}_\alpha - \fft{\zeta}}{\fft{x} \circ \conj{}(\fft{x}) + \mu}$.\quad\textbf{Solving} $\mathbf{w}^\star$:\quad $\mathbf{w} = \bm{\alpha} \circ \frac{\mu \mathbf{f}+\bm{\zeta}}{\mu+\frac{\gamma}{D}}$.

Here, $\circ$ denotes element-wise multiplication, and $\conj{}(\cdot)$ applies complex conjugate to the corresponding complex vector. We update the filter $\mathbf{w}_{t+1}$ as
\begin{equation}
    \mathbf{w}_{t+1}=(1-\eta_w)\mathbf{w}_t+\eta_w\tilde{\mathbf{w}},
    \label{eq:filter_w}
\end{equation}
where $\tilde{\mathbf{w}}$ is the result of the above optimization, $\eta_w$ is the filter learning rate. For updating other variables such as $\zeta$, we follow the standard practice from~\cite{lukezic2017discriminative,galoogahi2013multi}.

\noindent\textbf{Ensuring color model consistency.} We update the global color model as
\begin{equation}
    \mathbf{g}_t =(1-\eta_r)((1-\eta_g)\mathbf{g}_{t-1}+\eta_g \tilde{\mathbf{g}})+\eta_r\mathbf{g}_0.
\label{eq:color_model}
\end{equation}
Here, $\mathbf{g}_t=\{\mathbf{g}^f_t, \mathbf{g}^b_t\}$ is the color model for $t$-th tracing iteration. $\mathbf{g}_0$ is the starting color model extracted using the initial bounding box. $\tilde{\mathbf{g}}$ is estimated solely in current image slice. $\eta_g,\eta_r$ are the color model learning rate and retention rate, correspondingly. We include the retention rate to retain a certain portion of the initial states as the noisy labeling of the multi-spectral imaging technique unavoidably results in color drifts, i.e., a single neuron can be in different colors where different neurons with no spatial connections can be in the same color. In this case, we can make an early stop for our tracing process so that at least we can trust the current reconstructed trace. Therefore, we curb our tracer not going too far from the starting states, significantly decreasing the intractability of wrong traces. Note that, our color model can be easily generalized to single-channel fluorescent imaging modalities such as fMOST data~\cite{gong2016high} as in Results.

\subsection{Cross-section Determination Module}
\label{sec:cross_sec}
\noindent\textbf{Modeling cross-sectional plane of neuron.} A neuron is a 3D-volume structure such that if we can capture the cross-section of this 3D object every time, the structure of the tracing neuron can be progressively formed by the tiles of all the cross-sectional reconstructions. Intuitively, we take three planes every tracing iteration among corresponding $x$, $y$, and $z$ axes. Each time the neuron shows different projections in these three different planes ($xy,\,xz,\,zy$ plane). Here, we consider the plane with the most consistent appearance and shape as the cross-sectional plane and continue tracing the neuron, as in~\cref{fig:intro}. By tracing among the cross-sectional plane of the neuron, we maintain a relatively enough proportion of foreground to enable successful neuron tracing.

We assign a memory bank storing history of cross-plane reconstructions $\mathbf{S}=\{s_{n}\}_{n=1:i}$, where $s_{1}$ is the oldest reconstruction. For each iteration, we have different reconstructions for three planes $\{s^{xy}, s^{xz}, s^{zy}\}$, which are obtained by binarizing our alpha channel. The location of each plane window is calculated by spinning the current bounding box around its tracing center as a projection to each plane. To determine the cross-sectional plane, we first configure an average cross-section $\bar{s}$ among the memory bank $\mathbf{S}$ as,
\begin{equation}
    \bar{s}=\sum_{s\in \mathbf{S}}k(s;\sigma)\cdot s,
    \label{eq:mean_mask}
\end{equation}
where $k(\cdot)$ is a modified quadratic kernel, and $k(r,\sigma)=(1{-}(r/\sigma)^2)^2$ where $r$ is the distance of the iterative order of $\mathbf{S}$ to the last iteration and $\sigma$ is $\mathbf{S}$'s size. $\bar{s}$ represents height-width ratio ($\bar{s}_{hw}$) and area ($\bar{s}_{area}$) of the current bbox.

Next, we calculate height-to-width ratio and area for three reconstructions $\{s^{xy}, s^{xz}, s^{zy}\}$. The probability of each plane to be the cross-sectional plane is
\begin{equation}
    p=\sum_{s \in \{s_{hw},\,s_{area}\}}\lambda_c\cdot {\rm exp} \left(-\frac{(s-\bar{s})^2}{2\sigma_c^2} \right),
    \label{eq:cross_prob}
\end{equation}
where $\lambda_c$ is set to 0.5 to balance the effect of these two measurements, and $\sigma_c$ is set to be 0.4 in our system. We select the plane with the highest probability as the cross-sectional plane to be traced. By modeling within the history bank, we can adaptively enforce the tracing consistency.

\noindent\textbf{Modeling tracing direction.} Once the cross-sectional plane changes, we re-estimate the tracing direction (i.e., the next frame to be traced) for the new cross-sectional plane. This can done by taking out the latest $M$ tracing centers from the memory bank $S$ and setting the direction satisfying the tendency of those records. For example, if the new cross-sectional plane is $zy$ plane, we calculate the tendency of $x$ coordinates within the memory bank, and then navigate the tracing into the corresponding direction. 

\subsection{Bifurcation Determination Module}
\label{sec:branch}

\begin{figure}[!t]\small
  \centering
  \begin{minipage}[b]{0.4\textwidth}
    \includegraphics[width=\linewidth]{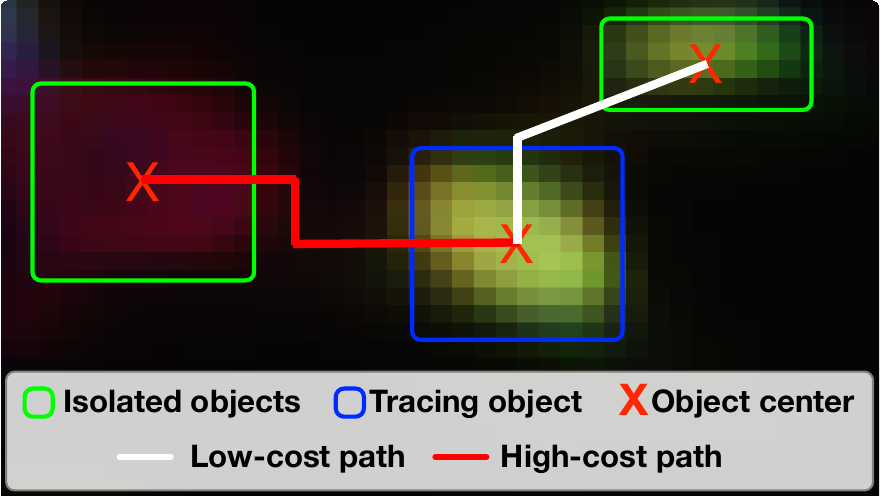}
    \captionsetup{skip=2pt,font=small}
    \caption{\textbf{An example of how to filter out objects for bifurcation}. Here, the one object with white path is retained while the one with red path is abandoned.}
    \label{fig:bifurcation_cand}
  \end{minipage}
  \hfill
  \begin{minipage}[b]{0.57\textwidth}
    \centering
    \includegraphics[width=0.87\linewidth]{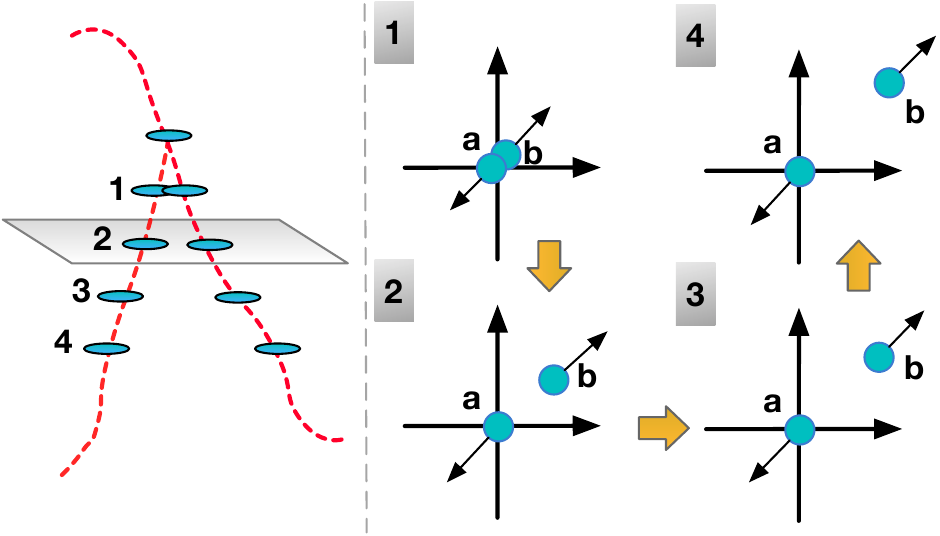}
    \captionsetup{skip=1pt,font=small}
    \caption{\textbf{Diagrammatic sketch of modeling bifurcation}.
    (\textbf{Left}) The overall trajectory (red line) in 3D.
    (\textbf{Right}) Detailed 2D movements.
    }
    \label{fig:bifurcation_sketch}
  \end{minipage}
\end{figure}

For each iteration, we reconstruct the neuron over a larger window around the tracing bbox as in Sec.~\ref{sec:aplha}. Within the resulting reconstruction, we acquire a set of isolated objects $O=\{o_i\}_{i=1:n}$, where $n$ is number of objects and $o_1$ is the current traced object. Next, we verify whether the reconstructed objects belong to the tracing neuron. For each pair of $\{o_1, o_i\}, i \in [1,n]$, using A$^\ast$~\cite{dechter1985generalized}, we find the optimal path $L$ from object $o_i$ to the current traced object $o_1$ with minimum cost. An example is shown in \cref{fig:bifurcation_cand}. Suppose $L=\{l_i\}_{i=1:N}$, where $N$ is the path length, and $l_i$ is a point on the path where $l_1, l_{N}$ are centrally located in $o_1, o_i$, respectively. Here, we define the cost as
\begin{equation}
    C=\sum_{i=1}^{N}\frac{\text{max}(\normd{l_i-l_1}, \normd{l_i-l_{N}})}{N}.
    \label{eq:pathcost}
\end{equation}
Here, the color distance of two points is normalized to $[0,1]$. This cost calculates the mean over the path for maximum Euclidean distances of each point from the starting and ending point. It is trivial to tell the range of the cost is [0, $c$] (see proof in Supplementary). After normalization for channels, we can obtain the probability $p_i$ from 0 to 1 by applying an exponential kernel as $p={\rm exp} \left(-\frac{C^2}{2c^2\sigma_b^2} \right)$, where $\sigma_b$ is the characteristic length-scale. Later, we examine whether the result probability $p_i$ is greater than $\tau_b$ and store the passed objects as candidates.

\noindent\textbf{Modeling bifurcation candidates.}
We consider bifurcation as the interaction of multiple objects. We use two objects as an example for demonstration, as in \cref{fig:bifurcation_sketch}. Considering the interaction of two objects, we assume that the bifurcation exists if the two objects are moving in face-to-face (FtF)/back-to-back (BtB) directions. Specifically, if we first observe only one object, and among the image stack, one object divides into two objects, and these two objects must move in BtB directions, within a reasonable period. A similar case goes to two objects converging into one object, where they should move in FtF directions. 
\section{Results}
\subsection{Experimental Setting}
\begin{SCfigure*}[1][!t]\small
    \centering
    \includegraphics[width=0.68\textwidth]{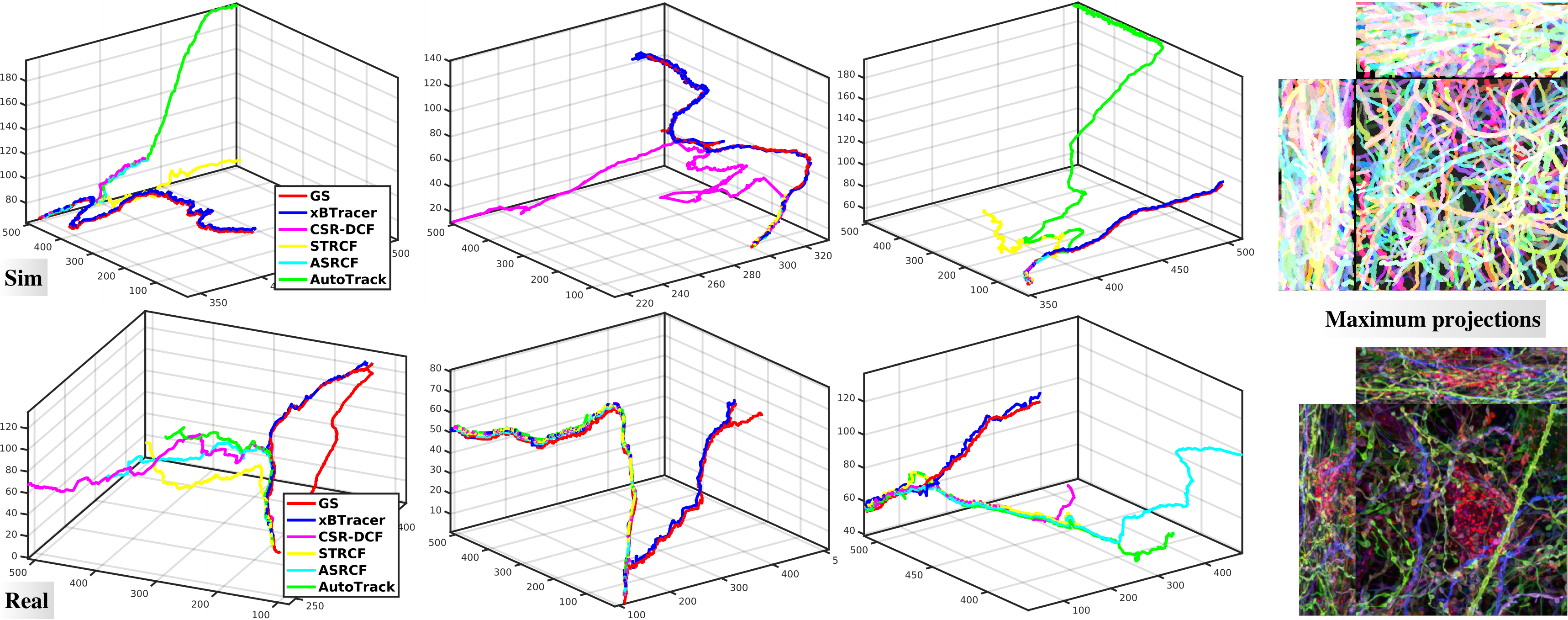}
    \caption{\textbf{Trace results comparisons.} \textbf{GS}: gold-standard. Our xBTracer perfectly fits the GS trace, while other tracers suffer severe problems.}
    \label{fig:sample_tracing}
\end{SCfigure*}
\noindent\textbf{Datasets.} We utilize two challenging datasets for evaluations.
\textbf{(i)~Simulated multi-spectral dataset.} To address this dataset shortage, we simulate four different but `realistic' multi-spectral stacks. We first dilate the human-labeled traces from real samples \cite{cai2013improved} and then use the simulation algorithm from~\cite{sumbul2016automated} to generate four stacks. These stacks have $512{\times}512{\times}200$ voxels, with different labeling densities about $8\%$, $15\%$, $20\%$, $30\%$, respectively.
\textbf{(ii)~Real-world dataset.} The real-world multi-spectral dataset is a confocal image from mouse hippocampal tissue~\cite{roossien2019multispectral}. It has $512{\times}512{\times}136$ voxels, and is imaged at 3 different spectral channels. The fluorescence density is $\sim30\%$, making it challenging even for human tracers. Sample frames for both datasets are in Supplementary.

\noindent\textbf{Evaluation metrics.} Four metrics are used to measure tracing quality. 
(\textbf{i})~Variation of Segment Alignment (\textbf{VSA}) score measures the agreement between the reconstructed trace and gold-standard (GS) trace. The range of the VSA score is [0, 1], where 0 means no pair of agreement, and 1 means perfect reconstruction. (\textbf{ii}) DIADEM~\cite{gillette2011diadem} is designed for evaluating the reconstruction of single neuron in a graph matching manner, range for 0 (worst) to 1 (best).(\textbf{iii})~Mean Fitting Distance (\textbf{MFD}) is specially designed to capture the distance of the reconstructed trace to the gold-standard trace, which reflects the out-of-track degree of tracers (\textbf{iv})~Bifurcation Retrieval Rate (\textbf{BRR})
reflects the ratio of bifurcations are recalled. BRR is only calculated when the GS trace has more than one branch. Details of the used metrics can be found in Supplementary.

\subsection{State-of-the-art Comparisons}
\noindent\textbf{Compared to object tracking methods.} 
We compare our method to different tracking methods, as reported in \cref{tab:trakcer} and \cref{fig:sample_tracing}. For the simulated dataset, compared with the most performed baseline, we have a 24\% performance gain, and an exciting improvement of 49\% for the most up-to-date work~\cite{li2020autotrack} in terms of VSA. For the real multi-spectral dataset, due to the limitations of current object trackers, we have a large performance gain in terms of all metrics, excelling all baseline methods. If we look into how close is the reconstructed trace to the gold-standard trace, traces generated by our tracer are most fitted as in~\cref{fig:sample_tracing}. As for bifurcations, since there are no mechanisms for object tracking methods to detect bifurcations, we reported them as 0. Meanwhile, we show that our designed modules can be applied to augment these methods, as in~\cref{tab:reb}, largely improving their performance when used for neuron tracing purpose.

\begin{table}[!t]\small
  \centering
  \adjustbox{valign=t}{\begin{minipage}{0.52\textwidth}
      \begin{minipage}{\linewidth}
        \centering\scalebox{0.7}{
        \begin{tabular}{ccccc}
            \toprule
            Dataset &  Method & VSA\textuparrow &MFD\textdownarrow& BRR\textuparrow\\
            \hline
            \scalebox{0.9}{\multirow{6}{*}{Sim.}}
            &CSR-DCF~\cite{lukezic2017discriminative}  &  {0.66} & \underline{33.47} & 0.00\\
            &STRCF~\cite{li2018learning}  & \underline{0.60} & 56.37 & 0.00\\
            &ASRCF~\cite{dai2019visual}  & 0.58 &  66.17 & 0.00\\
            &\textbf{xBTracer} (\textbf{Ours})&\textbf{0.82} & \textbf{23.71} & \textbf{0.18} \\
            \hline
            \scalebox{0.9}{\multirow{6}{*}{Real}}
            &CSR-DCF~\cite{lukezic2017discriminative} & {0.59} & \underline{35.68} & 0.00\\
            &STRCF~\cite{li2018learning} & \underline{0.56} & 37.12 & 0.00\\
            &ASRCF~\cite{dai2019visual} & 0.55 & 43.31 & 0.00\\
            &\textbf{xBTracer} (\textbf{Ours}) & \textbf{0.77} & \textbf{14.74} & \textbf{0.15}\\
            \bottomrule
        \end{tabular}}
        \captionsetup{skip=0pt,font=small}
        \caption{Compared to tracking methods. Except ours, others can not trace bifurcations.}
        \vspace{-0.1in}
        \label{tab:trakcer}
     \end{minipage}
     \begin{minipage}{\linewidth}
        \centering\scalebox{0.64}{
        \begin{tabular}{cccccc}
        \toprule
        Method & \cite{dai2019visual}/Aug.& \cite{li2018learning}/Aug.& \cite{li2020autotrack}/Aug.& \cite{lukezic2017discriminative}/Aug. & \cite{danelljan2017eco}/Aug. \\
        Note & CVPR19& CVPR18& CVPR20& CVPR17& CVPR17\\
        \midrule
        VSA\textbf{\textuparrow} & 0.55\,/\,0.72& 0.56\,/\,0.74 & 0.53\,/\,0.72& 0.59\,/\,\colorbox{blue!30}{0.77}& 0.54\,/\,0.73\\
        MFD\textbf{\textdownarrow} & 43.3\,/\,14.9& 37.1\,/\,\colorbox{blue!30}{14.6}& 46.5\,/\,14.8& 35.7\,/\,14.7& 40.8\,/\,14.8\\
        BRR\textbf{\textuparrow} & 0.00\,/\,0.14& 0.00\,/\,0.14& 0.00\,/\,0.13& 0.00\,/\,\colorbox{blue!30}{0.15}& 0.00\,/\,\colorbox{blue!30}{0.15}\\
        \bottomrule
        \end{tabular}}
        \captionsetup{skip=0pt,font=small}
        \caption{Augmenting other DCFs on real multi-spectral dataset.}
        \vspace{-0.1in}
        \label{tab:reb}
     \end{minipage}
     \begin{minipage}{\linewidth}
        \centering\scalebox{0.72}{
        \begin{tabular}{cccccc}
            \toprule
            Dataset &  Method & VSA\textuparrow &MFD\textdownarrow& BRR\textuparrow & DIADEM\textuparrow\\
            \hline
            \scalebox{0.9}{\multirow{3}{*}{Sim.}}& \cite{sumbul2016automated} & 0.72& \textbf{21.27}&\underline{0.08} & \textbf{0.91}\\
            & \cite{duan2021unsupervised}& \underline{0.74}& \underline{22.31} &0.05 & \textbf{0.91}\\
            &\textbf{Ours}& \textbf{0.82}& 23.71 &\textbf{0.18} & \underline{0.90}\\
            \hline
            \scalebox{0.9}{\multirow{3}{*}{Real}} & \cite{sumbul2016automated} & 0.61& \underline{14.09}&\underline{0.02} & \underline{0.84} \\
            & \cite{duan2021unsupervised}& \underline{0.63}& \textbf{13.53}& \underline{0.02} & \textbf{0.85}\\
            & \textbf{Ours}& \textbf{0.77}& 14.74& \textbf{0.15} & \underline{0.84}\\
            \bottomrule
        \end{tabular}}
        \captionsetup{skip=0pt,font=small}
        \caption{Compared to top multi-spectral neuron tracing methods.}
        \label{tab:tracing_methods}
    \end{minipage}
  \end{minipage}}
  \hfill
  \adjustbox{valign=t}{\begin{minipage}{0.45\textwidth}
    \begin{minipage}[t]{\linewidth}
    \centering\scalebox{0.65}{
    \begin{tabular}{ccccccccc}
        \toprule
         Metric & \cite{jin2019shutu}&\cite{huang2021automated} &\cite{januszewski2018high}& \cite{wang2019multiscale}& \cite{matejek2019biologically}& \cite{chen2021deep}&\cite{sheridan2023local}& \textbf{Ours}\\
        \midrule
        VSA${\uparrow}$ & 0.34& 0.38& 0.66& 0.65& 0.68& \underline{0.69}& \underline{0.69}&\textbf{0.77}\\
        MFD${\downarrow}$ & 15.87& 16.82& 14.84& 14.94& 14.77& \textbf{14.69}& 14.92& \underline{14.74}\\
        BRR${\uparrow}$  & 0.02& 0.02& 0.10& 0.08& 0.11& 0.09& \underline{0.12} &\textbf{0.15}\\
        DIADEM${\uparrow}$  & 0.82& 0.81& 0.80& 0.81& \underline{0.83}& 0.81& 0.82& \textbf{0.84}\\
        \bottomrule
    \end{tabular}}
    \captionsetup{skip=2pt,font=small}
    \caption{Compared to DL methods on real multi-spectral dataset.}
    \vspace{-0.02in}
    \label{tab:DL}
    \end{minipage}
    \begin{minipage}{\linewidth}
        \includegraphics[width=\linewidth]{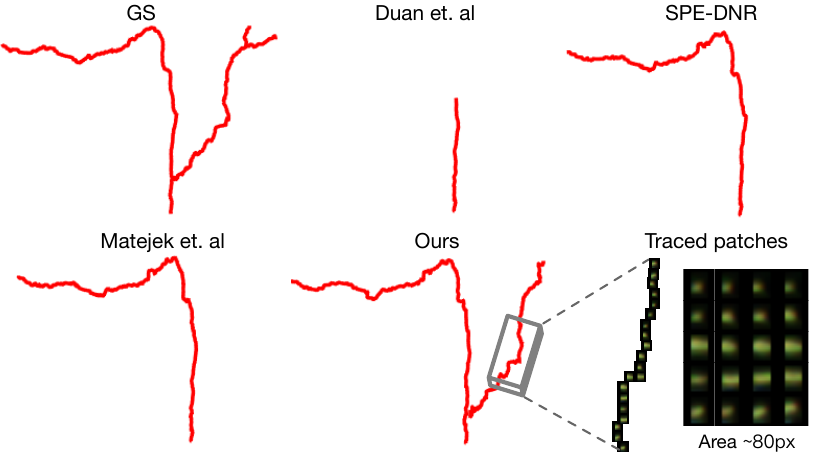}
        \captionsetup{skip=0pt,font=small}
        \captionof{figure}{\textbf{Comparisons to the most performing neuron tracing methods.} We can observe more compact reconstruction using our method. The average area of the traced bboxes is $\sim$80 pixels, where we pad our traced patches to the same size for better visualization on the bottom right.}
        \label{fig:comp}
    \end{minipage}
  \end{minipage}}
\end{table}
\noindent\textbf{Compared to tracing methods developed for multi-spectral images.} The results are shown in \cref{tab:tracing_methods}, where we also report DIADEM scores for the partially reconstructed traces. We find that \cite{sumbul2016automated,duan2021unsupervised} can suffer from fragmented segmentation problems, leading to broken traces and thus lower VSA scores. Compared to these two methods, while our tracer maintains a similar level of MFD and DIADEM score, it generates more consistent traces to the GS traces.

\noindent\textbf{Compared to deep learning (DL) methods developed for other imaging modalities.} Since there are no DL methods developed for multi-spectral images, we adapt methods from other modalities, where the models are trained using 25\% of GS traces (line-like annotated), after dilation of 5 voxels in each dimension, as annotations. Then, neurons are separated based on spatial connectivity. The results are shown in~\cref{tab:DL}, where the trace results compared to the most performed methods are in~\cref{fig:comp}. For~\cite{jin2019shutu,huang2021automated}, we find that their methods mostly focus on single neuron reconstruction which has much sparser labeling density, so is hard to generalize to the densely-labeled multi-spectral images. For~\cite{januszewski2018high,chen2021deep,wang2019multiscale,matejek2019biologically,sheridan2023local}, these offline-trained models are still not comparable to our proposed training-free method. We can observe that our method produces more complete traces than others (\cref{fig:comp}), resulting in a higher VSA score. Overall, bifurcation is the most challenging part and requires a lot of work. We hope that our new methodology can inspire more future work.
\begin{table}[!t]
    \adjustbox{valign=t}{\begin{minipage}{0.52\textwidth}
        \setlength{\tabcolsep}{5pt}
        \centering\scalebox{0.82}{
        \begin{tabular}{clccc}
            \toprule
            Dataset &  Setting & VSA\textuparrow & MFD\textdownarrow & BRR\textuparrow\\
            \hline
            \scalebox{0.9}{\multirow{4}{*}{Sim.}}
            & w/o\_c &  0.71 & 27.38 & 0.12\\
            & w/o\_b & 0.76 & 27.22 & 0.00\\
            & w/o\_m & 0.73 &  29.71 & 0.11\\
            & full model & \textbf{0.82} & \textbf{23.71} & \textbf{0.18} \\
            \hline
            \scalebox{0.9}{\multirow{4}{*}{Real}}
            & w/o\_c &  0.70 & 24.21 & 0.10\\
            & w/o\_b & 0.72 & 36.40 & 0.00\\
            & w/o\_m & 0.66 &  34.29 & 0.09\\
            & full model & \textbf{0.77} & \textbf{14.74} & \textbf{0.15} \\
            \bottomrule
        \end{tabular}}
        \captionsetup{skip=0pt,font=small}
        \caption{Ablative study of our method.}
        \label{tab:ablation}
    \end{minipage}}
    \hfill
    \adjustbox{valign=t}{\begin{minipage}{0.44\textwidth}
        \centering
        \includegraphics[width=\linewidth]{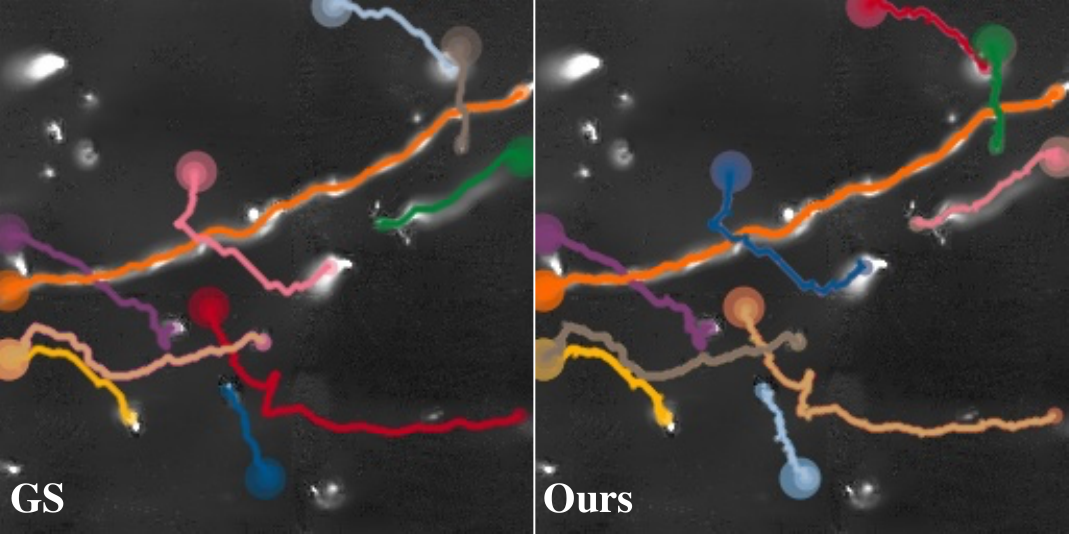}
        \captionsetup{skip=0pt,font=small}
        \captionof{figure}{Applying our methodology to fMOST~\cite{yang2019fmst} imaging modality.}
        \label{fig:fmost}
    \end{minipage}}
    \vspace{-0.2in}
\end{table}

\subsection{Ablative Study} 
To validate our components, we derive four variants, i.e., \textbf{(i) w/o\_c} stands for the variant without cross-section determination module; \textbf{(ii) w/o\_b} as a variant with the removal of bifurcation module; \textbf{(iii) w/o\_m} denotes that we only use naive DCF rather than our enhanced DCF; \textbf{(iv) full model} is the method with all designed modules. Ablation results are reported in \cref{tab:ablation}. Overall, removing the alpha mask constraint downgrades the most performance. The reason is that our tracing is mostly based on reconstruction such that a more compact reconstruction can lead to a smooth and successful trace. It is interesting to find that our incomplete variants still outperform the other DCF tracking or neuron tracing baselines at most times, which further verifies the effectiveness of our proposed algorithm. Overall, each module contributes to keeping our tracer on track, which is important for neuron tracing as generally, we cannot use out-of-track traces. We can also observe that our traces are more centered, meaning our full model is more robust for the difficult multi-spectral images. We also test the generalization ability of our method in other imaging modalities such as fMOST~\cite{yang2019fmst} data, as in~\cref{fig:fmost}, where we observe high agreements over traces.
\section{Conclusions}
In this work, we propose online multi-spectral neuron tracing, which is a novel and different methodology for tracing neurons in multi-spectral images. Our method is easy to set up and training-free, where the superior performance compared to different methods, both object tracking and neuron tracing methods, certifies its effectiveness. Besides, we test our method on different imaging modalities, namely multi-spectral and fMOST data, both showing promising qualitative and quantitative results, so that one can use our method to generate fast and accurate neuron reconstructions. We intend to investigate more complicated nervous systems with the proposed tracer in our future work.

{\small
\bibliographystyle{ieee_fullname}
\bibliography{egbib}
}

\end{document}